\documentclass[fleqn,10pt]{SelfArx} 

\definecolor{color1}{RGB}{0,0,90} 
\definecolor{color2}{RGB}{0,20,20} 

\usepackage{hyperref} 
\hypersetup{hidelinks,colorlinks,breaklinks=true,urlcolor=color2,citecolor=color1,linkcolor=color1,bookmarksopen=false,pdftitle={Title},pdfauthor={Author},urlcolor=blue}

\usepackage{graphicx}
\usepackage{natbib}
\usepackage{amsmath}
\usepackage{multirow}
\usepackage{subfigure}

\newcommand{\n}[1]{\mathrm{#1}}

\JournalInfo{Published in Journal of the American Ceramic Society, Vol. 98 (11), 3490-3495, 2015} 
\Archive{\href{http://dx.doi.org/10.1111/jace.13701}{DOI: 10.1111/jace.13701}} 

\PaperTitle{Modeling the microstructural evolution during constrained sintering} 

\Authors{R. Bj\o{}rk, H. L. Frandsen and N. Pryds} 
\affiliation{\textit{Department of Energy Conversion and Storage, Technical University of Denmark - DTU, Frederiksborgvej 399, DK-4000 Roskilde, Denmark}} 
\affiliation{*\textbf{Corresponding author}: rabj@dtu.dk} 

\Keywords{} 
\newcommand{\keywordname}{Keywords} 

\Abstract{A numerical model able to simulate solid-state constrained sintering is presented. The model couples an existing kinetic Monte Carlo (kMC) model for free sintering with a finite element model (FEM) for calculating stresses on a microstructural level. The microstructural response to the local stress as well as the FEM calculation of the stress field from the microstructural evolution is discussed. The sintering behavior of a sample constrained by a rigid substrate is simulated. The constrained sintering results in a larger number of pores near the substrate, as well as anisotropic sintering shrinkage, with significantly enhanced strain in the central upper part of the sample surface, and minimal strain at the edges near the substrate. All these features have also previously been observed experimentally.}

\begin{document}

\flushbottom 

\maketitle 


\thispagestyle{empty} 

\section{Introduction}
Many applications require components that are sintered under a variety of constraining stresses. These constraints may be internal, such as those that arise during co-sintering of multiple layers densifying at different rates. Or the constraints may be external due to a rigid substrate constraining the densification of the adjacent surface or be due to an applied load.

A general review of constrained sintering, discussing both continuum and microstructural evolution, is presented by Green et. al. \cite{Green_2008}. Experimentally, samples sintering under constraints display anisotropic macroscopic shrinkage and porosity evolution. The overall change in size of a component sintering under constrains can be modelled macroscopically, but linking the macroscopic anisotropic shrinkage to anisotropy in the microstructure is challenging \cite{Green_2008}.

In this work, we present a numerical model able to model solid-state constrained sintering of a microstructure. Furthermore, this model is applied to study the microstructural evolution of a sample constrained to sinter on a rigid substrate, as this is an experimentally well characterized system. Constrained sintering of both porous gold and silver circuit film on a rigid substrate has been studied experimentally \cite{Choe_1995, Lin_2004}. The constrained films had higher porosity, when compared to freely sintered films, but no change in grain size was observed. Tape cast and dip-coated alumina films sintered on a rigid substrate had pores perpendicular to the substrate surface and a higher porosity near the substrate \cite{Guillon_2007a,Guillon_2007b}. Elongated pores have also been observed for thin zirconia films sintered on rigid substrates \cite{Mucke_2009}. Synchrotron computed microtomography has been used to study constrained viscous sintering of glass on a substrate \cite{Bernard_2011}, and several other studied on constrained viscous sintering of glass exist \cite{Calata_1998,Ollagnier_2009,Ollagnier_2010}.


A numerical model that is able to simulate constrained solid-state sintering and predict microstructural properties, such as pore size and structure, would be of interest for a number of technological applications as macroscopic properties can depend directly on the microstructural parameters \cite{Boonyongmaneerat_2007}. A number of such numerical models, capable of modeling constrained sintering on the mesoscale currently exist. For early stage sintering, discrete element method (DEM) models have replicated experimental observed trends, such as large and orientated pores near the substrate \cite{Martin_2009,Rasp_2012}. A surface-evolver program has also been used to calculate the equilibrium configuration of particles during constrained sintering \cite{Wakai_2003, Wakai_2012}, but this approach is significantly computationally intensive. A finite element model (FEM) approach using diffusion equations for lattice and surface diffusion \cite{Bruchon_2012} has also been used to model constrained sintering, but in this model grain growth was not modeled and the number of modeled grains were less than 100. Finally, a coupled Monte Carlo and FEM approach has also been used to simulate sintering of two particles, with the energy defined partly from mass flow in the system and partly from elastic strain energy caused by node displacements in the finite element model \cite{Bordere_2002}. A coupled kinetic Monte Carlo (kMC) and FEM approach has also been used to model multi-scale sintering \cite{Olevsky_2006,Molla_2014}, where the kMC model was used to derive local material properties, such as the normalized shear and bulk viscosities, which were then utilized in the macro-scale FEM model. However, no coupling from the macro-scale FEM model back to the meso-scale kMC model was presented, and thus the microstructure did not evolve according to any constraining conditions.

In this work, we present a mesoscale model capable of simulating the entire sintering process under constraining conditions. The model extends an existing kinetic Monte Carlo numerical model, which has previously been shown to correctly model the free sintering of regular structures \cite{Olevsky_2005}, copper spheres \cite{Tikare_2010, Cardona_2011}, close-packed spheres \cite{Bjoerk_2012b} and powder with a given particle size distribution \cite{Bjoerk_2012a}. This model is coupled with a FEM to allow for the calculation of microstructural stresses during sintering. These local stresses are incorporated in the kMC model to simulate densification due to both curvature and local stresses. Using a microstructural model to study constrained sintering is an ideal approach for studying the evolution of a microstructure as an initial powder can be sintered with and without constraints, allowing for a direct comparison of the results.

The coupling between the kMC and the FEM must go both ways; i.e. the microstructure (in the kMC model) must evolve in response to a stress field, and the stress field must be calculated due to microstructural events. Coupling a kMC model with a FEM for materials processes has previously been used for studying recrystallization in metals, where the FEM calculates deformation of a sample, while the Monte Carlo model simulate grain growth \cite{Sarma_1996, Radhakrishnan_1998, Raabe_2000, Wang_2009}. Using voxel data obtained from e.g. a kMC sintering model, a method has also been developed that uses a finite element approach to determine e.g. tensile stress-strain curves due to failure and damage \cite{Mishnaevsky_2005}. A viscoplastic finite element continuum model, coupled with a Monte Carlo model for calculating microstructural strain rates have also been presented \cite{Mori_2004}. The work presented here differs from these approaches in that the active sintering process itself, which is annihilation of vacancies at grain boundaries, is modified according to the stresses obtained using the FEM model. In contrary to the Monte Carlo simulation of grain growth, the current annihilation events changes the microstructure, so remeshing of the FEM model is required.

\section{The kinetic Monte Carlo model of sintering}
The kinetic Monte Carlo model for free solid-state sintering is described in detail elsewhere \cite{Tikare_2010, Cardona_2011}. In the model, individual grains and pores are defined on a two or three dimensional square/cubical grid. Here we generally refer to a single grid cell as a voxel in both two and three dimensions. The model simulates grain growth, pore migration and vacancy formation and annihilation through diffusion processes. The driving force for sintering in the model is the capillarity (reduction of interfacial free energy), which is defined by the neighbor interaction energy between voxels. The energy, $E$, at a given site, $i$, is proportional to the sum of unlike neighbors to that site, i.e.
\begin{eqnarray}\label{Eq.Def_energy}
E_i = \frac{1}{2}\sum_{j=1}^{8\;\mathrm{or}\;26}J*(1-\delta{}(q_i,q_j))~,
\end{eqnarray}
where $J$ is the neighbor interaction energy, $q_i$ and $q_j$ are the state of the sites $i$ and $j$ respectively, and $\delta$ is the Kronecker delta function. A value of $J=1$ is chosen, as this constitutes the simplest case possible. Grains are identified by different values of $q$, while all pore sites have the same state, $q=0$. The number of neighboring sites considered is 8 for a two dimensional simulation and 26 for a three dimensional one.

Sintering is modeled by interchanging two neighboring sites, altering the state of a single grain site or collapsing an isolated pore site, called a vacancy, by moving it to the surface of the sample. If any of these ''movements`` lowers the total energy of the system, as calculated by Eq. (\ref{Eq.Def_energy}), then the move is accepted, whereas if the energy is increased the move may be accepted with probability, $P$, based on the standard Metropolis algorithm, defined as

\begin{eqnarray}\label{Eq.Def_prop}
P = \left\{ \begin{array}{ll}
\mathrm{exp}\left(\frac{-\Delta{}E}{k_\n{B}T}\right) & \textrm{for } \Delta{}E > 0\\
1 & \textrm{for } \Delta{}E \leq{} 0\\
\end{array} \right.
\end{eqnarray}

where $T$ is the temperature, $k_\n{B}$ is Boltzmann's constant and where the different types of events can have different temperatures. The attempt frequency of each type of event can also be varied, allowing different magnitudes of pore surface diffusion, grain boundary diffusion and grain boundary mobility. The attempt frequency is the probability that a given type of move is attempted. The kMC model is currently dimensionless. Time is measured in Monte Carlo steps which are proportional to real time \cite{Tikare_2010}. Under constrained sintering, the stresses present are an additional driving force for sintering, and a stress will cause a change in the probability that a given kMC move is accepted or rejected.

\section{Including stress in the kinetic Monte Carlo model}
To model constrained sintering, the evolution of the microstructure must depend on the local stress. In the stress-free kMC model, sintering happens purely in response to the topology of the local environment, i.e. the surrounding voxel sites, but when stresses are included in the model the microstructure must also respond to the global environment of the sample. To do so a FEM has been implemented and coupled to the existing kMC model to allow the stress to be calculated based on events that causes strain in the microstructure. The FEM allows for a direct calculation of the stress field throughout a sample based on the mechanical properties of the solid part of the sample backbone, i.e. Young's modulus and Poisson's ratio. The FEM is written in C++ and the direct solution of the FEM is computed using Eigen v.3 \cite{eigenweb}.

The stress must be calculated on at least the scale at which the microstructural evolution takes place, i.e. at the scale of the voxel. As sub-voxel resolution is not meaningful, the computational mesh for the FEM is chosen to resolve the microstructure at the voxel level. This is a sufficient resolution, as the FEM will not be used to capture surface stresses, as the energy due to surface tension, i.e. sintering stresses, is inherently modeled in the kMC model through minimization of surface energy \cite{Cardona_2012}.

In two dimensions, a single voxel can be meshed either as two linear triangular elements or as one bilinear element. In three dimensions, the choice is between five linear tetrahedral elements or one trilinear element. Regardless of choice, the number of nodes in the mesh remains the same, but the connectivity of the nodes is different. The simplest mesh is chosen, which is the bilinear and trilinear elements for the two- and three dimensional cases, respectively. Thus, each voxel in the kMC model corresponds to a single finite element. Bilinear element are superior compared to trilinear elements, in elastic problems \cite{Brauer_1993,Benzley_1995}, although for computational reasons only linear and not higher order elements are chosen here. The pore sites, which are empty sites in the kMC model, are not meshed. The microstructure is remeshed completely for each new Monte Carlo timestep, as the morphology changes.

In the kMC model, the active processes during solid-state sintering, i.e. grain growth, pore migration by surface diffusion, and vacancy generation, diffusion and annihilation are the processes that could be influenced by the local stress. Experiments on constrained sintering have shown that grain growth during sintering is not significantly affected by the stress field \cite{Choe_1995,Guillon_2007b}, and hence grain growth in the model is not modified according to the local stress. In the model presented here, pore migration moves are not dependent on the local stress, but the effect of stress on pore migration will be discussed in a future work. This means that only the kMC annihilation process, which is also the only process resulting in densification, i.e. sintering, is modified to depend on the local stress.

\subsection{Annihilation events under stress}
The developed model is a phenomenological model that modifies both the probability and direction of annihilation to depend on the compliance (``deformability'') of the microstructure at the place, where the annihilation takes place. Experimental observations indicate that samples under compression densify more rapidly, because the work performed by the external forces contributes to overcome the activation energy.

\subsubsection{Stress evaluation}
\begin{figure}[!t]
  \centering
  \includegraphics[width=0.5\textwidth]{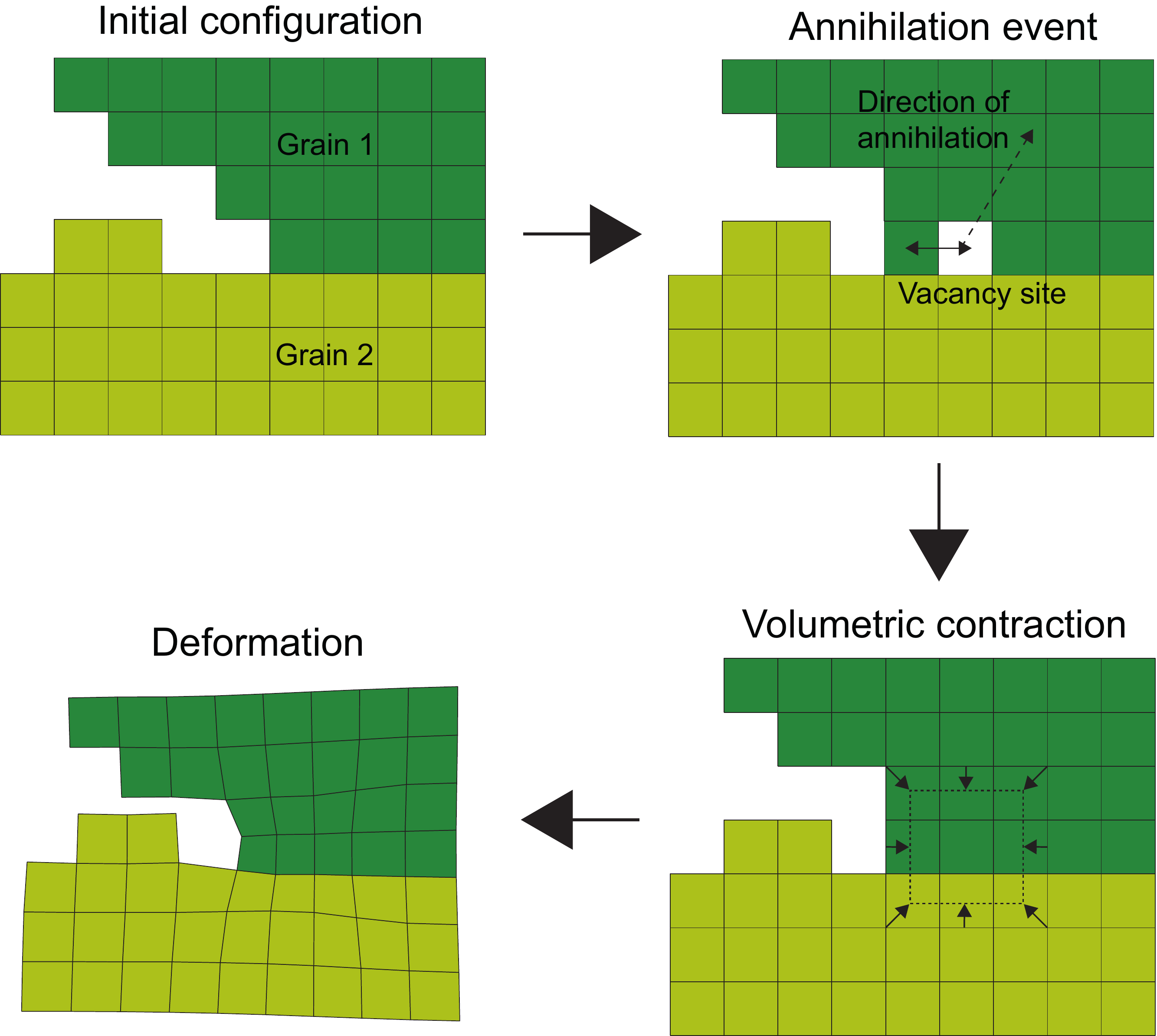}
  \caption{An illustration of the stress generating process. A single vacancy site is generated at the grain boundary between two grains and annihilated. This corresponds to a volumetric strain, shown here in the surrounding sites, which causes a deformation of the microstructure.}
  \label{Fig_Annihilation_illustration}
\end{figure}

To simulate the stress during an annihilation a fixed volumetric contraction is applied in the FEM at the site, where the annihilation event is occurring and its surrounding sites, as shown in Fig. \ref{Fig_Annihilation_illustration}. On the kMC meso-scale, the annihilation step is an instantaneous event in time, and thus stresses will be generated immediately. For this reason the FEM is a purely elastic model. However, the macro-scale shear and bulk moduli can still be derived from the microstructure \cite{Molla_2014}. The computed stress will be used to calculate a probability of annihilation, as will be discussed subsequently.

After the elastic response determined in the FEM calculation, the volumetric strain will subsequently relax over time depending on the viscosity of the material. In the current model this relaxation is assumed to relax all stress before the next Monte Carlo time step. The exact influence of this is something for future investigations.

The mesh for the FEM calculation is only re-established once for every kMC step, in order to reduce computation time. The actual FEM stress calculations are however done for each individual annihilation event.

The model presented here bear resemblance to the inelastic mechanical model introduced by Rice \cite{Rice_1971}. In this model, changes in internal variables can cause a system to respond elastically. Here, the strains in the FEM result from changes in internal variables of the system, namely the annihilation event occurring in the kMC microstructure. These annihilations can be viewed as inelastic deformations, which then cause the remaining system to respond elastically through the FEM.

\subsubsection{Annihilation probability}
\begin{figure}[!t]
  \centering
  \includegraphics[width=0.5\textwidth]{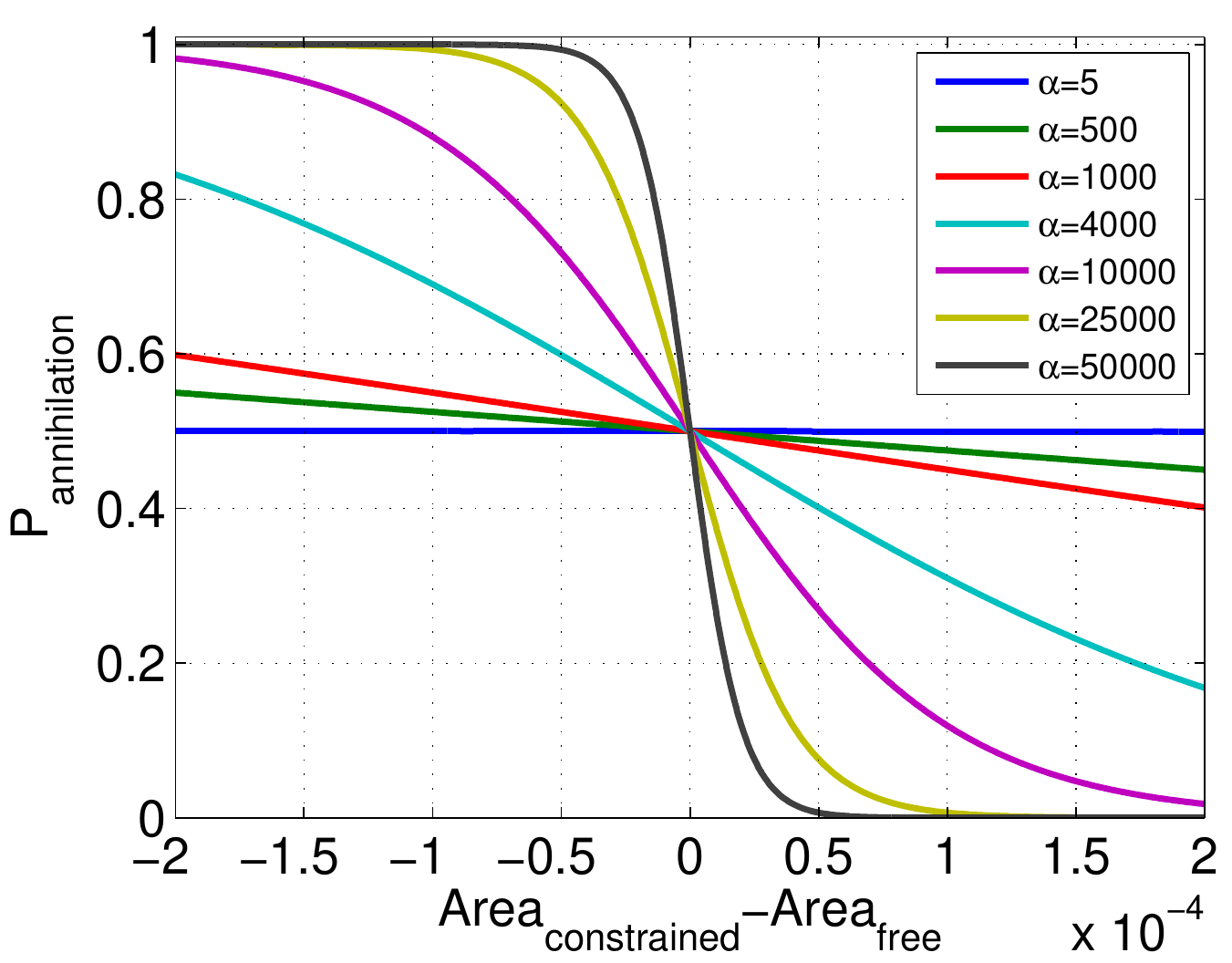}
  \caption{The probability to annihilate a vacancy site, depending on the compliance of the microstructure in a 2D case. A negative difference occurs for sites under compression, for which the sintering probability is increased. As the FEM is elastic, this could equally well have been the difference in stress due to a volumetric contraction.}
  \label{Fig_Annihilation_frequency}
\end{figure}

Since the FEM is fully elastic, annihilation is equivalent to appling a volumetric force and consider the change in volume in Fig. \ref{Fig_Annihilation_illustration}. The change in volume of the voxels caused by this volumetric contraction is then evaluated in two cases: free sintering, and when the body is constrained. The difference in contraction volume is a direct measure of the extra stresses occurring because of the constriction. In terms of volumetric force, the measure is how compliant the local microstructure is to densifications at that point in the sample. For simplicity, a volumetric force is considered subsequently. Here, the difference in volume, $V$, of the voxels between free and constrained sintering is weighted by a tangent hyperbolic function, to calculate an additional probability, $P_\n{a}$, that the vacancy will annihilate, as
\begin{eqnarray}\label{Eq_P_modified}
P_\n{a} = \frac{1}{2}\left(1-\n{tanh}\left(\alpha\left(\n{V}_\n{constrained}-\n{V}_\n{free}\right)\right)\right).
\end{eqnarray}
The constant $\alpha$ determines the degree of coupling between the stress field and the probability of annihilation and thus the microstructural evolution, as shown in Fig. \ref{Fig_Annihilation_frequency}. Thus, in this model, regions under tension will densify slower or not at all, i.e. it will be harder for the surrounding sites to fill the vacancy formed and to densify the material. Regions under compression will sinter more rapidly. Sites that are freely sintering will only have half the probability to annihilate as compared to the free kMC model, but this is simply remedied by doubling the attempt frequency of annihilation events.

The value of the coupling constant, $\alpha$, is influenced by the chosen applied volumetric force and the value of Young's modulus. The wide range of the $\alpha$-parameter, shown in Fig. \ref{Fig_Annihilation_frequency}, is caused by the choice of Young's modulus and volumetric contraction force, as these directly influence the change in volume.

The probability in Eq. (\ref{Eq_P_modified}) is only evaluated for vacancy annihilation moves that is already accepted in the kMC annihilation algorithm, i.e. those that fulfill the probability calculated in Eq. (\ref{Eq.Def_prop}). This is because it has previously been reported that the evolution in surface energy during constrained sintering is approximately 1000 times greater than that of the internal strain energy \cite{Bordere_2002}. This means that the overall vacancy annihilation is driven by the kMC diffusion and the FEM calculated probability is only a minor perturbation of the kMC probability.

\subsubsection{Direction of annihilation}
\begin{figure}[!t]
  \centering
  \includegraphics[width=0.5\textwidth]{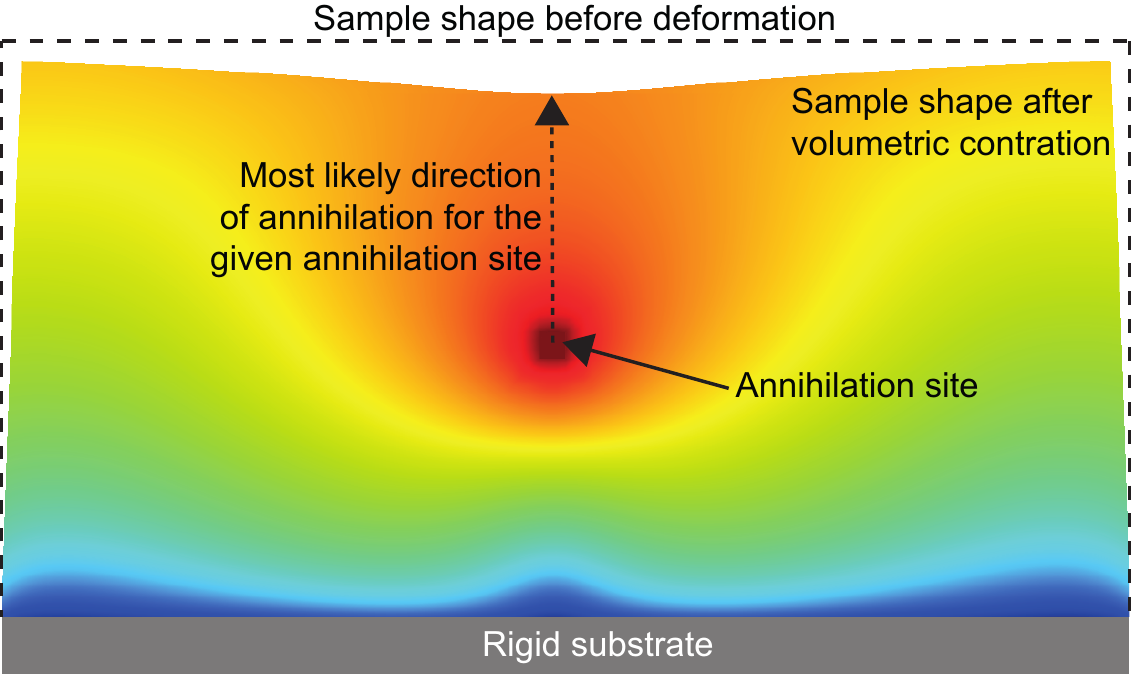}
  \caption{An illustration of the computations to find the direction of annihilation under constrained sintering. The image shows the scaled deformation of a sample constrained on a rigid substrate, when the indicated annihilation site is subjected to a volumetric contraction force. The color indicate the logarithm of the solid displacement. The mean strain is computed on each of the sample's four faces. The most probable direction of annihilation is at the greatest strain, as indicated.}
  \label{Fig_Annihilation_direction}
\end{figure}

In a successful annihilation move, the isolated vacancy site is annihilated by collapsing a column of sites from the vacancy to the surface of the sample into the vacancy. The direction chosen from the vacancy site to the sample surface must, in free sintering, be distributed such that an equal number of annihilations terminate at all surfaces of the sample \cite{Bjoerk_2014a}. When constrained sintering is modelled, the direction of annihilation will depend on the constraining conditions. The direction of annihilation will not be uniform, but must depend on the global shape deformation of the sample due to the local densification, with the preferred direction being the direction of the largest strain, as illustrated in Fig. \ref{Fig_Annihilation_direction}.

To reflect this behavior, a new algorithm was developed. In this algorithm, the strain at the surface of the sample is calculated and used to select the annihilation direction. The presented algorithm is developed for a rectangular sample, but could be generalized according to a sample of arbitrary shape \cite{Bjoerk_2014a}. The algorithm is as follows: First, the whole sample is meshed, with a single mesh element for every 10x10 voxels, i.e. ten times as rough as the mesh used for the compliance calculation discussed above. All pores are filled in this model, as the strain at the sample surface will not depend on small pores located throughout the sample. The 10:1 ratio of the mesh is used partly for computational reasons and partly in order to avoid determining the surface of the sample at the level of the individual voxel, as this is notoriously complicated.

A volumetric force is then applied at the position of the annihilating voxel, and the mean strain at the surface of the sample in then determined at each of the faces of the sample. The same volumetric force as for the compliance model discussed above is used when the rough mesh is applied. However, this is not of importance, as it is the normalized strain that is determined. The mean strain is then compared with that of the other faces and subsequently used as the probability that the annihilation path will terminate at that face, such that the annihilation path is most likely to terminate at the face with the largest mean strain. Once a face has been selected, the strain at all sites on that face is then normalized and used to select the end point for the annihilation path. The scheme is illustrated in Fig. \ref{Fig_Annihilation_direction} and in bullets, the algorithm is
\flushleft
\begin{enumerate}
\begin{itemize}
\item Remove pores and remesh sample in 10:1 ratio
\item Apply volumetric contraction at annihilating voxel
\item Compute mean strain on each sample face
\item Select face for annihilation direction based on mean strain
\item Select point on face based on strain at each voxel on selected face
\end{itemize}
\end{enumerate}

Splitting the probability into two factors, i.e. first selecting a face and then a point on that face, ensures that surfaces with more voxels are not favored for annihilation, as this will result in an erroneous strain of the sample \cite{Bjoerk_2014a}.

\section{Sintering on a rigid substrate - A case study}
In order to verify the model, constrained sintering of a sample on a rigid substrate is modelled, as this is experimentally well investigated. As discussed previously, a sample sintering constrained on a rigid substrate displays three characteristics; anisotropic shrinkage, more pores closer to the substrate and elongated pores perpendicular to the substrate. The goal of the work presented here is to investigate if the model reproduce some of these features. The results of the modeling simulation will not be directly compared to an experimental data set, as the non-dimensionality of the kMC model does not easily allow for this. The purpose of this study is to demonstrate that the developed model provides a qualitatively correct result. A two dimensional sample is considered, for computational reasons. For the presented case simulated with the kMC/FEM model the total simulation time increases about a factor of 5 for the two dimensional simulations considered here, as compared to free sintering with the kMC model only.

\subsection{Model setup}
\begin{figure*}[p]
  \begin{center}
  \subfigure[Initial microstructure]{\includegraphics[width=0.7\textwidth]{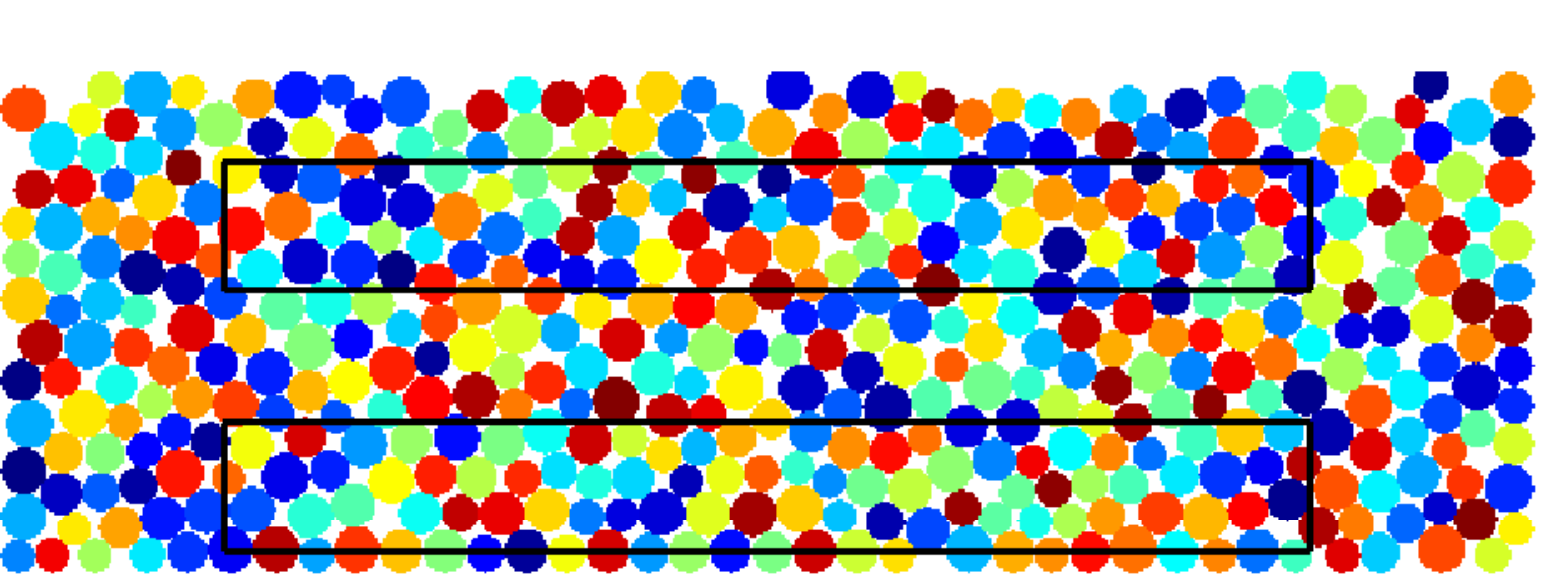}}\hspace{0.1cm}
  \subfigure[Weak coupling, 500,000 MONTE CARLO STEPS, $\alpha = 5$]{\includegraphics[width=0.7\textwidth]{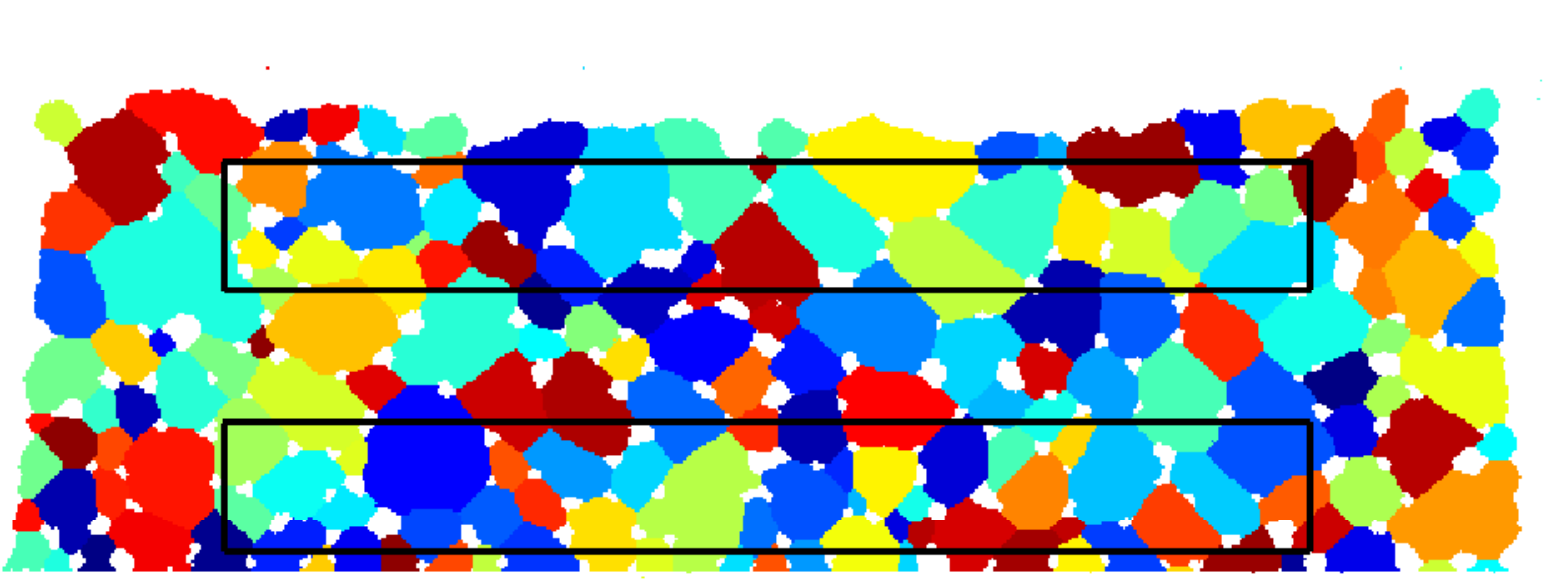}}\hspace{0.1cm}
  \subfigure[Strong coupling, 500,000 MONTE CARLO STEPS, $\alpha=50000$]{\includegraphics[width=0.7\textwidth]{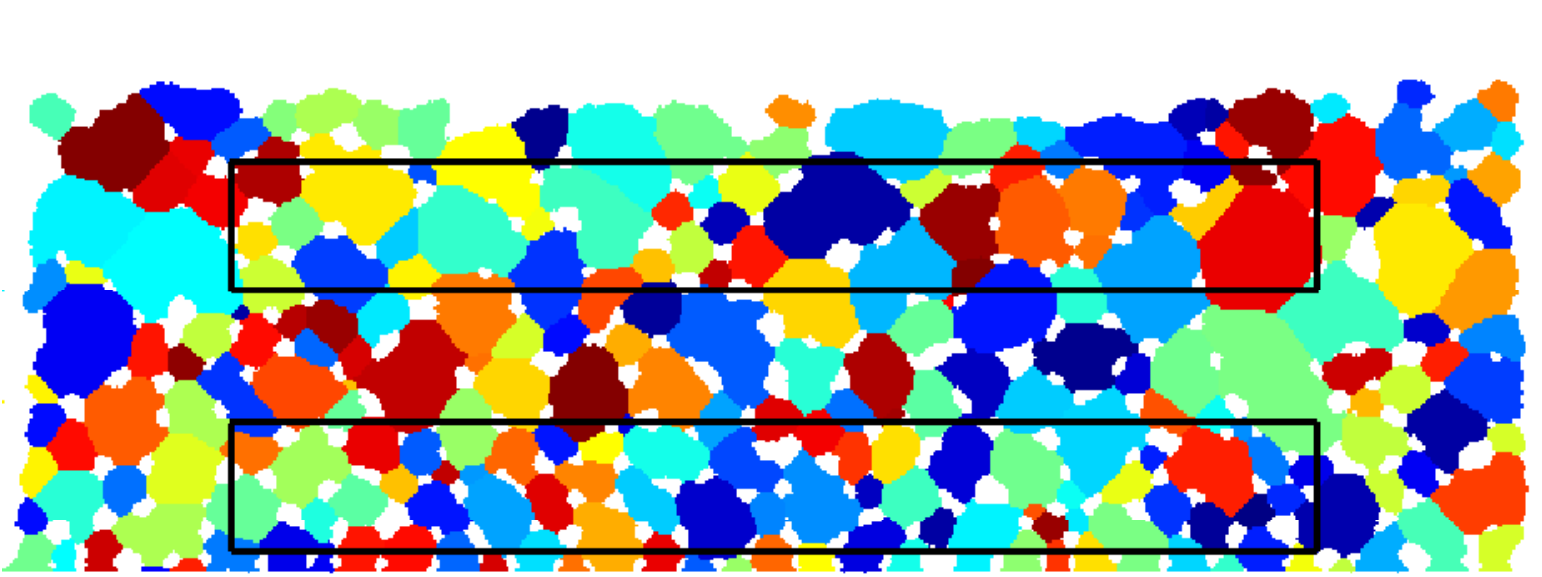}}
  \end{center}
  \caption{The initial microstructure (a) and the microstructure after 500,000 Monte Carlo steps for a sample sintering constrained to a rigid substrate and with weak (b) or a strong (c) coupling between the stress and the microstructure. The black boxes shows the areas where the relative density was characterized.}
  \label{Fig_Microstructure}
\end{figure*}

The two dimensional sample considered consists of randomly packed spherical particles, with diameters uniformly distributed between 13 and 19 pixels. The sample was generated by simulating the pouring of these particles into a container using the Large-scale Atomic/Molecular Massively Parallel Simulator (LAMMPS) code. The sample dimensions were 600x200 pixels. The initial sample geometry is shown in Fig. \ref{Fig_Microstructure}. The kMC event temperatures are taken to be 0, 0.7 and 1 for grain growth, pore surface diffusion, and vacancy annihilation respectively. The attempt frequencies are taken to be 0.5, 5 and 2.4 for the same mechanisms, respectively. These values allow sufficient pore migration to even out the pores between annihilation events. Poisson's ratio is chosen to be 0.3, which is a typical value for most materials studied in sintering. Young's modulus is taken to be 1 Pa, and the applied volumetric force to be 0.1 N. These latter two values influence only the range of $\alpha$, and do not affect the simulation otherwise.

The sintering of the sample was simulated using the above mentioned numerical framework, with the lower surface of the sample constrained to a rigid surface.

\subsection{Influence of the coupling variable}
The resulting microstructure after 500,000 Monte Carlo steps of the sample sintered on a rigid substrate with both a very weak and a strong coupling between the stress field and the annihilation event probability, $\alpha = 5$ and $50000$, respectively, is shown in Fig. \ref{Fig_Microstructure}. It can clearly be seen that the constrained sample with the strong coupling has more pores near the substrate. The stress field generated by the constraining surface implemented in the model, through the coupling between stress and microstructure, results in this behavior.

\begin{figure}[!ht]
  \centering
  \includegraphics[width=0.5\textwidth]{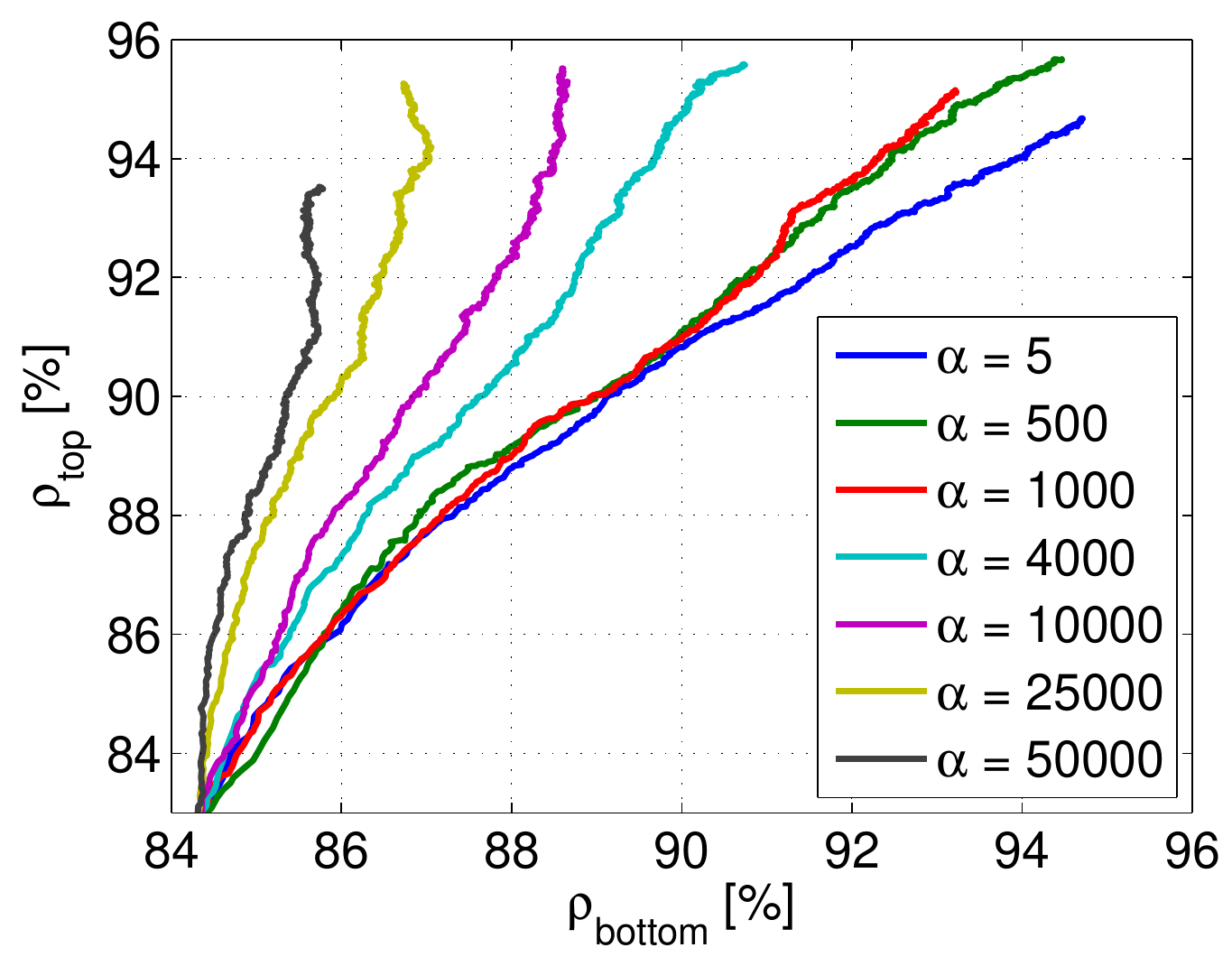}
  \caption{The relative density in the upper and lower parts of the microstructure (as indicated on Fig. \ref{Fig_Microstructure}) for various coupling strengths. As time evolves and the sample sinters, the relative density increases. The high initial relative density is caused by the two dimensional sample.}
  \label{Fig_Pore_frac_slices_reduced}
\end{figure}

The influence of the coupling between the event probability and the stress field is best illustrated by analyzing the distribution of pores in different parts of the sample, as function of the coupling parameter, $\alpha$. This is illustrated in Fig. \ref{Fig_Pore_frac_slices_reduced}, where the relative density in the upper and lower part of the sample, for various coupling parameters, is shown. The upper and lower parts are those indicated in Fig. \ref{Fig_Microstructure}. As can be seen from Fig. \ref{Fig_Pore_frac_slices_reduced}, a weak coupling results in a behavior equivalent to free sintering, with equal density in the different parts of the sample, while a strong coupling significantly reduces the sintering evolution in the part of the sample closest to the constraining surface, resulting in higher porosity in this region.

An example of the strong or weak coupling between the microstructure and the constraining conditions is e.g. the experimental study of Mohanram et. al. \cite{Mohanram_2006}, which can be seen as an analogue to the above numerical experiment. Here two ceramic-filled glass systems were fired under constraining conditions. One glass crystallized towards the end of sintering, with an increase of elastic modulus as a consequence. This resulted in a strong coupling to the stress field, which led to a in-homogeneous distribution of pores at the end of sintering. The other glass remained viscous throughout the sintering, why a homogeneous distribution of pores was observed, which corresponds to the weak coupling.

\subsection{Sample shape after sintering}
Sintering of samples constrained on a rigid substrate is known to result in more contraction in the direction perpendicular to the substrate, compared to the direction parallel to the substrate \cite{Guillon_2007a}. Development of a curved surface, with greater contraction near the top of the sample has also been observed \cite{Rasp_2012}.

\begin{figure}[t]
  \centering
  \includegraphics[width=0.5\textwidth]{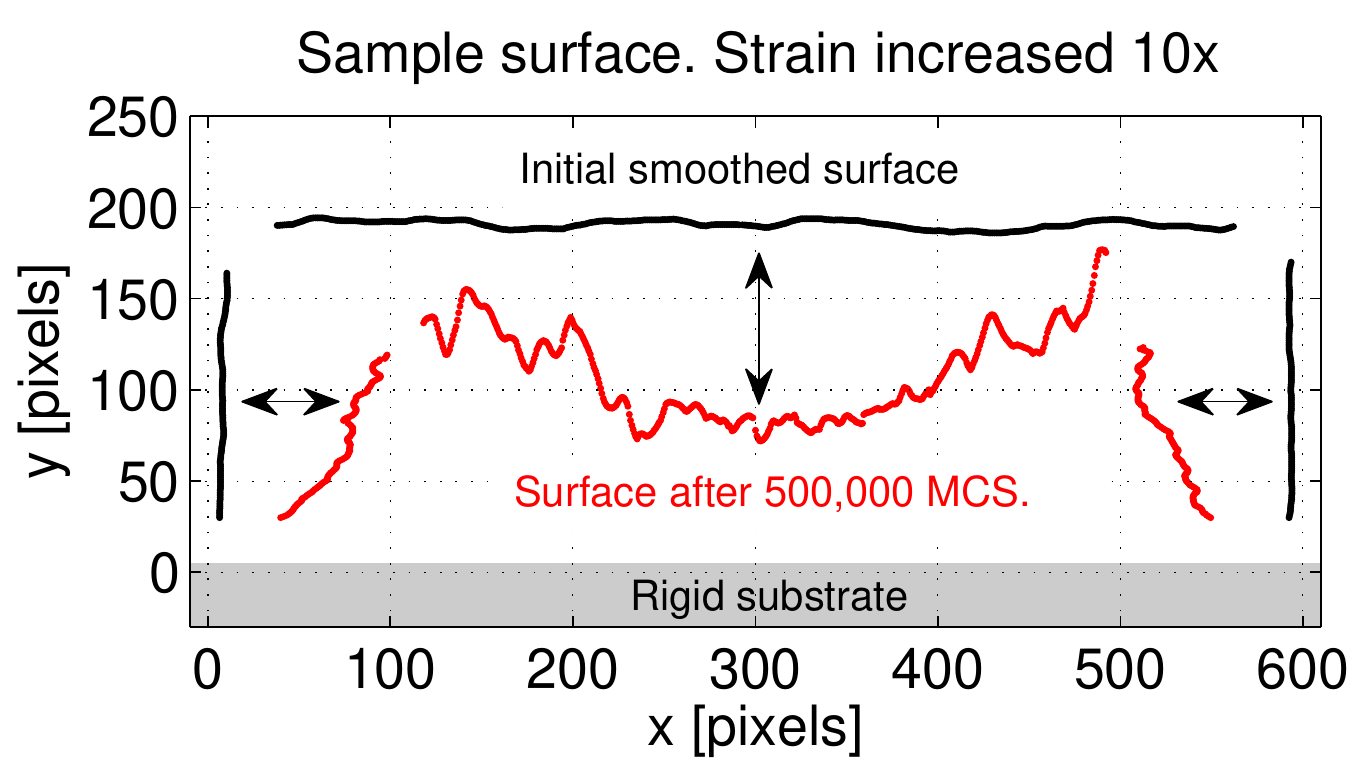}
  \caption{The initial surface of the sample and after 500,000 Monte Carlo steps for the case of $\alpha = 50000$. The strain has been enhanced by a factor of ten, for visualization. The surface is smoothed through a running average 50 pixels wide, and therefore the surface is not computed near corners.}
  \label{Fig_Strain_surface_t_500_alpha_50000}
\end{figure}

The surface of the sample for $\alpha=50000$ is shown in Fig. \ref{Fig_Strain_surface_t_500_alpha_50000} before and after sintering. To even out the roughness of the surface, caused by the individual voxels, the surface is a computed averaged over the surface at 25 pixels on either side of a given point. This also means that the surface is not determined at the outmost corners of the sample. As can be seen from the figure, the surface clearly shrinks more in the direction perpendicular to the substrate at the center of the sample. In addition, the contraction is greater on the vertical sides far from the substrate than closer to the substrate.

\begin{figure}[!h]
  \centering
  \includegraphics[width=0.5\textwidth]{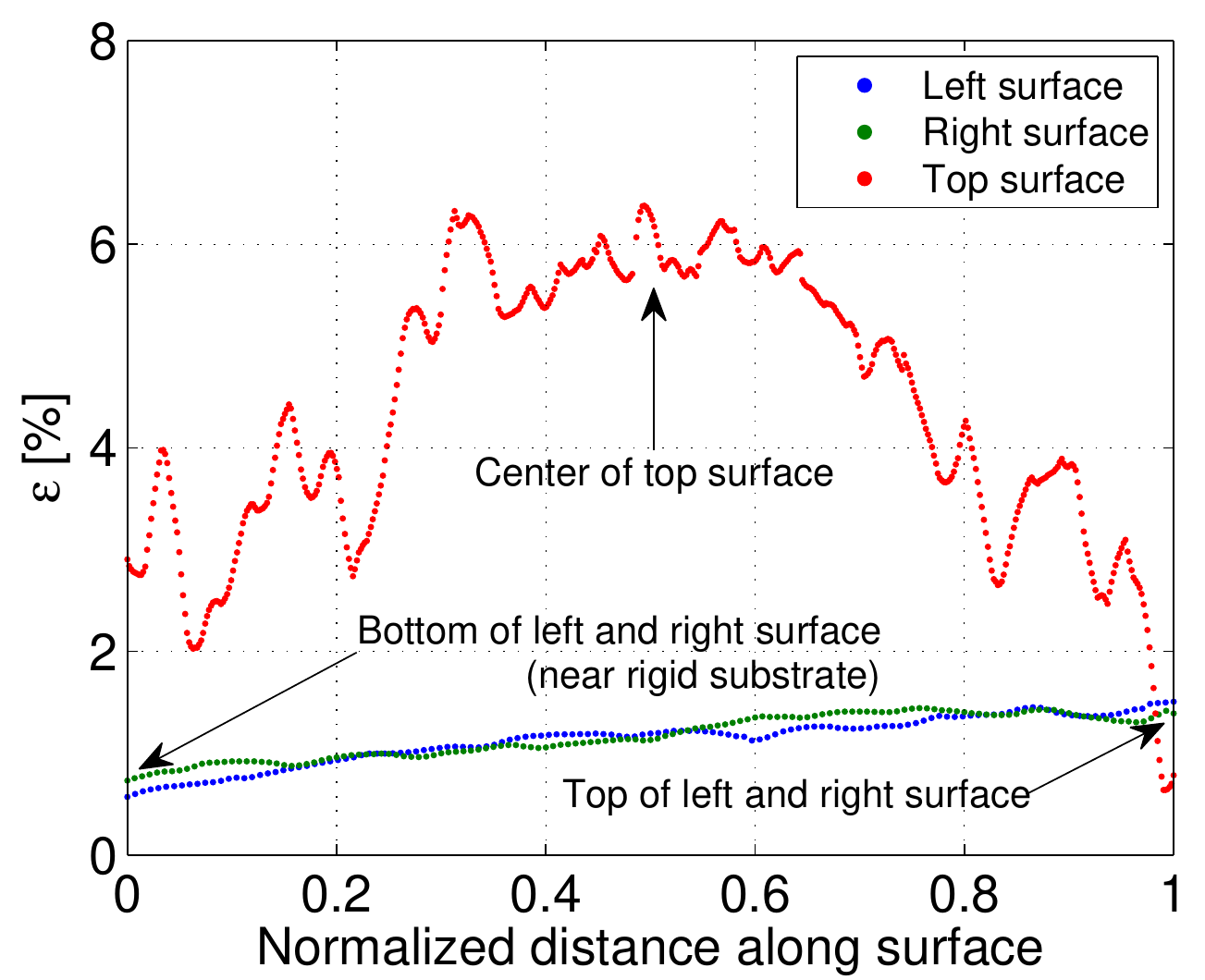}
  \caption{The strain at different points on the smoothed surface after 500,000 Monte Carlo steps for the case of $\alpha = 50000$. The corners are excluded, i.e. it is only along the part of the surface shown in Fig. \ref{Fig_Strain_surface_t_500_alpha_50000}.}
  \label{Fig_Strain_constrained_t_500_alpha_50000}
\end{figure}

The strain at each point on the smoothed surface is also shown in Fig. \ref{Fig_Strain_constrained_t_500_alpha_50000}, where the strain can be seen to be a factor of three higher in the center of the top surface, compared to close to the corners. The evolution of the shape of the sample agrees qualitatively with experimentally observed shaped deformations \cite{Guillon_2007a,Rasp_2012}, where a greatly enhanced shrinkage of the top surface relative to the left and right vertical surfaces of a sample sintering on a rigid substrate has been observed.

\section{Conclusion}
The microstructural response to a stress field and the derivation of the stress field from the microstructural evolution has been discussed. A new microstructural model to describe constrained sintering has been developed. This couples a kMC sintering model and a FEM for calculating stresses because of constraints and annihilations during sintering. As a case study, the sintering of a sample on a rigid substrate was investigated and the effect of the coupling between the stress field and the microstructure was shown to result in a larger number of pores near the substrate. Anisotropic strain of the sample was also observed, with significantly enhanced contraction in the center of the top sample surface, and minimal strain on the edges near the substrate. Both increased number of pores and the observed anisotropic strain were also previously observed experimentally.

\section*{Acknowledgements}
The authors would like to thank the Danish Council for Independent Research Technology and Production Sciences (FTP) which is part of The Danish Agency for Science, Technology and Innovation (FI) (Project \# 09-072888) for sponsoring the OPTIMAC research work.  The authors which to thank Dr. Veena Tikare who developed the used kMC model, as well as Dr. Michael Braginsky, whose unpublished work inspired some of the approaches presented here.


\end{document}